\documentclass[12pt,a4paper]{article}
\usepackage{graphicx}
\begin{document}
 
\title{\bf Towards identifying the world stock market cross-correlations: \\
DAX versus Dow Jones}
 
\author{S. Dro\.zd\.z$^{1,2}$, F. Gr\"ummer${^1}$, F. Ruf$^{3}$ 
and J. Speth$^{1}$ \\
\\
$^{1}$Institut f\"ur Kernphysik, Forschungszentrum J\"ulich,\\
D-52425 J\"ulich, Germany \\
$^{2}$ Institute of Nuclear Physics, PL-31-342 Krak\'ow, Poland\\
$^{3}$ WestLB International S.A., 32-34 bd Grande-Duch. Charl.,\\
L-2014 Luxembourg}
\date{\today}
\maketitle

\begin{abstract}
Effects connected with the world globalization 
affect also the financial markets.
On a way towards quantifying the related characteristics we study
the financial empirical correlation matrix of the 60 companies which both
the Deutsche Aktienindex (DAX) and the Dow Jones (DJ) industrial average
comprised during the years 1990-1999. The time-dependence of the underlying
cross-correlations is monitored using a time window of 60 trading days.
Our study shows that if the time-zone delays are properly
accounted for the two distant markets largely merge into one. 
This effect is particularly visible during the last few years.
It is however the Dow Jones which dictates the trend.

\end{abstract}
 
 
\bigskip
 
\newpage

\section{Introduction}

The fundamental feature of the time-evolution of self-organizing
complex dynamical systems, such as financial markets, is a permanent
coexistence and competition between noise and collectivity.
Noise is ubiquitous and overwhelming,
and therefore it seems natural that majority of eigenvalues of the stock
market correlation matrix agree very well~\cite{Lalo,Pler}
with the universal predictions
of random matrix theory~\cite{Meht}.
This perhaps can be traced back to similar characteristics observed already
on the level of human's brain activity~\cite{Kwap}
Collectivity on the other hand is much more subtle but it is this component
which is of principal interest because it accounts for system-specific
non-random properties and thus potentially encodes the system's future states.

In the correlation matrix formalism collectivity can be attributed 
to deviations from the random matrix predictions.
Our related recent study~\cite{Droz} based on both, the Deutsche Aktienindex
(DAX) and the Dow Jones (DJ) industrial average
point to a nontrivial time-dependence of the resulting
correlations. As a rule, the drawdowns are found to be always accompanied
by a sizable separation of one strong collective eigenstate 
of the correlation matrix which, at the same time, 
reduces the variance of the noise states. 
The drawups, on the other hand, turn out to be more competitive.
In this case the dynamics spreads more uniformly over the eigenstates 
of the correlation matrix.

All the above mentioned results are based on studies of the single stock
markets, in isolation to all the others. 
An every day experience indicates, however, an increasing role of effects
connected with the world globalization, which seems also to affect the
financial markets.
It is therefore of great interest to quantify the related characteristics.
Besides their significance for understanding the mechanism of evolution
of the contemporary stock markets they may also be relevant for practical
aspects of the theory of optimal portfolios and risk management~\cite{Mark}.
On a way towards exploring this issue below we study the cross-correlations
between all the stocks comprised  by DAX and by Dow Jones.
Both include the same number (30) of the companies and in space terms
represent two distant and at the same time leading markets in their area.
Mixing them up results in 60 companies which determines the size of the
correlation matrix to be studied.

\section{DAX + DJ correlation matrix}

In general for an assets labelled with $i$ and represented
by the price time-series $x_i(t)$ of length $T$ 
one defines a parallel time-series of normalized returns

\begin{equation}
g_i(t) = {{G_i(t) - \langle G_i(t) \rangle_t} \over v^2},
\label{eq:g}
\end{equation}
where

\begin{equation}
G_i(t) = \ln x_i(t+\tau) - \ln x_i(t),
\label{eq:Gg}
\end{equation}
are unnormalized returns,

\begin{equation}
v = \sigma(G_i) = 
\sqrt {\langle G_i^2(t) \rangle_t - \langle G_i(t) \rangle_t^2}
\label{eq:v}
\end{equation}
is the volatility of $G_i(t)$ and $\tau$ denotes the time lag imposed.
For $N$ stocks the corresponding time-series $g_i(t)$ of length $T$ 
are then used to form
a $N \times T$ rectangular matrix $\bf M$. 
Then, the correlation matrix is defined 
as 
\begin{equation}
{\bf C} = {1\over T} {\bf M} {\bf \tilde M}.
\label{eq:C}
\end{equation}
where the tilde denotes the transposed matrix.
In our specific case of the two stock markets 
the matrix ${\bf M}$ is formed from the time-series of both
the DAX  $(g_i^{DAX}(t))$ and the Dow Jones $(g_j^{DJ}(t))$ normalised returns.
Then the corresponding global $({\cal G})$
correlation matrix ${\bf C}_{\cal G}$ can be considered
to have the following block structure:
\begin{eqnarray}
{\bf C}_{\cal G} = 
\left ( \matrix{ {\bf C}_{DAX,DAX} & {\bf C}_{DAX,DJ} \cr
                 {\bf C}_{DJ,DAX}  & {\bf C}_{DJ,DJ} } \right)
\label{eq:CG}
\end{eqnarray}

As our previous study shows~\cite{Droz} 
the dynamics of the stock market correlations
shows a very nontrivial time-dependence. In order to detect such effects
the preferred time window $T$ is to be as small as possible. However,
in order not to artificially reduce the rank of the correlation matrix,
and thus in order not to introduce any spurious collectivity, $T$ needs to
be at least equal to $N$. This sets the lowest limit on a time window which
can be used to study the time-dependence of correlations.   
In the present case of incorporating all the stocks traded by DAX and by 
Dow Jones based on the daily price changes ($\tau=1$ trading day)
this corresponds to $T=60$ trading days.
The total time-interval explored here covers the years 1990-1999.

\section{Results}

To begin with when inspecting the nature of correlations it
is instructive to look at the distribution of matrix elements of ${\bf C}$.
For ${\bf C}_{\cal G}$ such a distribution obtained by averaging over all the
$T=60$ time windows is displayed by the solid line in Fig.~1.
Contrary to the single stock market case it visibly deviates from a pure
Gaussian-like shape. Decomposing this distribution into its components
originating from the internal correlations between the DAX 
$({\bf C}_{DAX,DAX})$, the Dow Jones $({\bf C}_{DJ,DJ})$ and from the
cross-correlations between the two (${\bf C}_{DAX,DJ}$ and ${\bf C}_{DJ,DAX}$),
explains the global structure. All the individual distributions are 
Gaussian-shaped but centered at different locations.
Consistent with our previous study~\cite{Droz} which points to stronger
collectivity effects in DAX relative to Dow Jones the distribution
associated with DAX is shifted more to positive values as compared to 
the Dow Jones.
The distribution of matrix elements connecting these two stock markets 
is centered at a value which is much closer to zero.
This can be interpreted in terms of a significantly
weaker strength and more random character of such cross-correlations
than the ones inside DAX and Dow Jones respectively.

The above observations remain in accord with the following more detailed
study in terms of the time-dependence of eigenspectrum
$\{ \lambda^i \}$ of ${\bf C}_{\cal G}$ calculated with the time step
of one trading day over the time interval $T$ of the past 60 trading
days.
Such a time-dependence of the four largest eigenvalues versus the corresponding
two indices (DAX and DJ) is illustrated in Fig.~2.
In contrast to a single stock market case where dynamics is typically
dominated by one outlying eigenvalue here one can systematically 
identify the two large eigenvalues. The range of variation of the remaining 
eigenvalues is on average compatible with the limits prescribed~\cite{Edel} by
entirely random correlations:
\begin{equation}
\lambda^{max}_{min} = \sigma^2 (1 + 1/Q \pm 2\sqrt{1/Q}),
\label{eq:Q}
\end{equation}
where $Q=T/N$ and $\sigma^2$ equals to the variance of the time series.
In our case both these quantities equal unity which results in
$\lambda^{max}=4$.

In fact the two largest eigenvalues represent the two stock markets
as if they were largely independent. Comparing the time-dependences
of the largest eigenvalue $\lambda_{\cal G}^1$ of ${\bf C}_{\cal G}$
with the largest eigenvalue $\lambda_{DAX}^1$ of ${\bf C}_{DAX,DAX}$ 
and the second largest eigenvalue $\lambda_{\cal G}^2$ of ${\bf C}_{\cal G}$
with the largest eigenvalue $\lambda_{DJ}^1$ of ${\bf C}_{DJ,DJ}$,
as is shown in Fig.~3, clearly points to such a conclusion.
In formal terms
the structure of eigenspectrum thus indicates that the two submatrices,
${\bf C}_{DAX,DAX}$ and ${\bf C}_{DJ,DJ}$,
remain largely disconnected and this in fact is compatible
with the (not far from zero centered) Gaussian distribution of the
connecting matrix elements of ${\bf C}_{DAX,DJ}$ and ${\bf C}_{DJ,DAX}$.
At the same time however $\lambda_{\cal G}^1$ and $\lambda_{\cal G}^2$
(similarly as $\lambda_{DAX}^1$ and $\lambda_{DJ}^1$) go in parallel
as far as their time-dependence is concerned, especially over the last few
years.
This in turn signals sizable correlations between them which in fact
seems natural because the DAX and the Dow Jones increases and decreases
respectively display
significant correlations in time as can be seen from Fig.~2.

From the technical perspective reconciling these somewhat confliciting 
conclusions turns out more straightforward than expected 
and at the same time leads to a new very interesting result.
By constructing the correlation matrix ${\bf C}_{\cal G}$ from $g_j^{DJ}(t)$
and $g_i^{DAX}(t+1)$, i.e., the DAX returns are taken one day advanced
relative to the Dow Jones returns,
one obtains the eigenspectrum whose structure significantly changes.
Its resulting time-dependence is shown in Fig.~4. Now, except for the early
90's, one large eigenvalue dominates the dynamics which means that a sort
of a one common market emerges. 
Consistently, it also obeys the characteristics observed
before~\cite{Droz} for the single markets: as a rule the collectivity
of the dynamics is weaker (smaller $\lambda^1_{\cal G}$) during increases than
during decreases.
The origin of such a significant change of the eigenspectrum when going
from the situation of Fig.~2 to the one of Fig.~4 is of course associated
with the matrix elements connecting the two markets.
As shown in Fig.~5 their distribution is now more asymmetric relative to zero
which leads to an amplified coupling between $C_{DAX,DAX}$ and $C_{DJ,DJ}$.

The significance of this result can be appreciated when looking (Fig.~6) at the
eigenspectrum of the correlation matrix where time-ordering between DAX
and Dow Jones is interchanged, i.e.,
${\bf C}_{\cal G}$ is built up from $g_j^{DJ}(t+1)$
and $g_i^{DAX}(t)$. The reminders of correlations that can still be
claimed to be present in the case described by Fig.~2
are now seen to be completely washed out
and the two sectors become almost entirely uncorrelated.
This fact is also globally expressed by the distribution of the corresponding
matrix elements shown in Fig.~7. Those connecting DAX and Dow Jones are now
centered basically at zero.

\section{Summary}

In summary, the present study of the time-dependence of cross-correlations
between all the stocks comprised by DAX and by Dow Jones
points to a significant novel element associated with dynamics
of the contemporary financial evolution.
By properly taking into account the time-zone delays both these markets
largely merge into a single one. This becomes particularly spectacular
in the last few years. At the same time an emerging global market preserves
the distinct difference in the mechanism governing increases and decreases,
respectively. Similarly as for single markets~\cite{Droz} the increases
also in this case are less collective and more competitive than decreases.
This study also documents that it is the Dow Jones which takes a leading
role in this emerging global market. As an interesting problem for further
study it seems likely that such a global world market involves many
other markets as well.

\newpage

\begin{figure}[ht]
\begin{center}
\leavevmode
\includegraphics[angle=270,width=15cm]{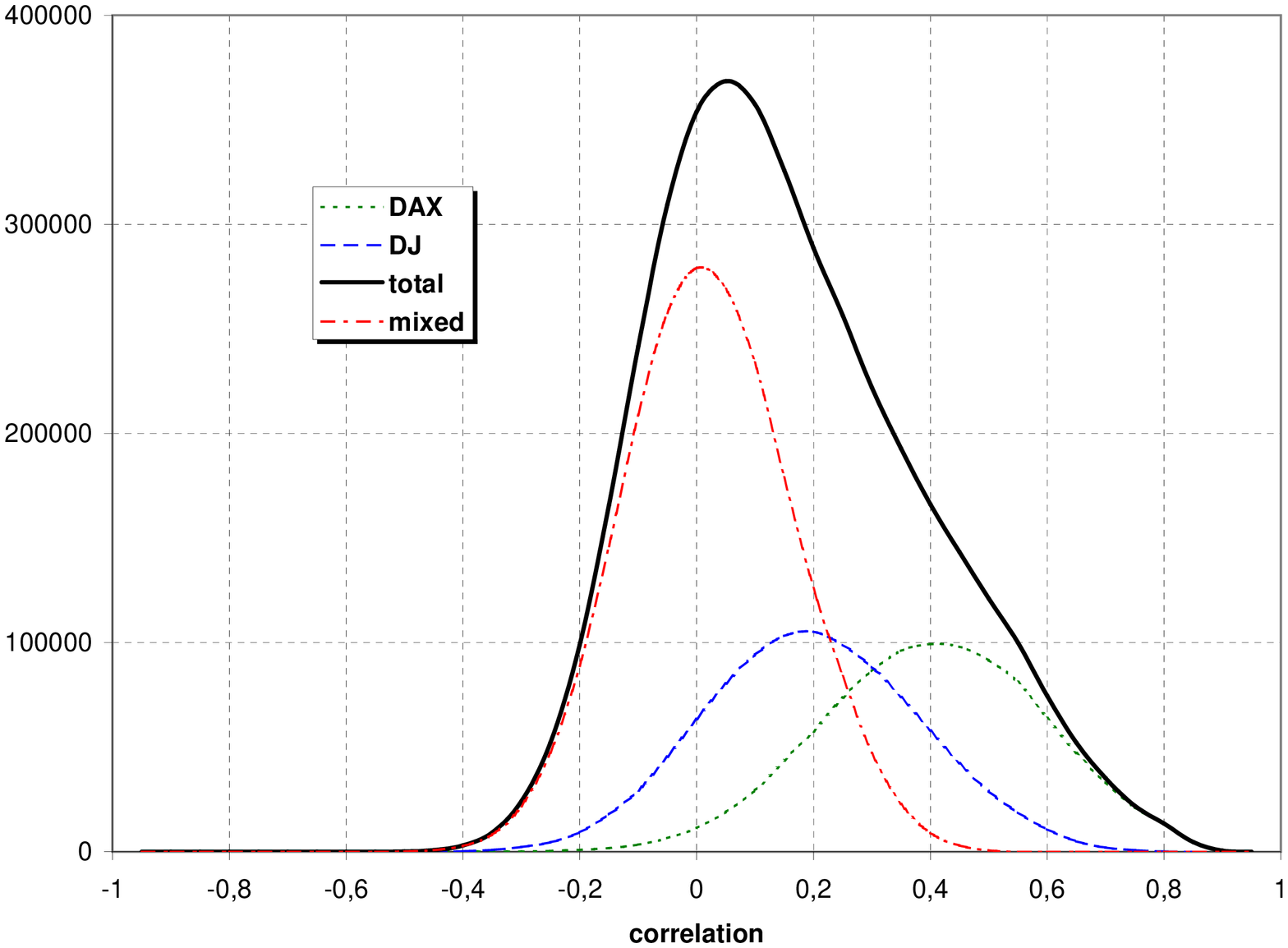}
\end{center}
\caption{\it Distribution of matrix elements of the
correlation matrix ${\bf C}_{\cal G}$ (solid line) calculated from
the daily price variation of all $N=60$ companies comprised by
DAX and by Dow Jones. The individual contributions originating from
${\bf C}_{DAX,DAX}$ (dotted line), ${\bf C}_{DJ,DJ}$ (dashed line)
and from the connecting matrix elements of ${\bf C}_{DAX,DJ}$ and
${\bf C}_{DJ,DAX}$ (dashed-dotted line) are also shown. These distributions
are obtained by averaging over all the $T=60$ trading day time windows
covering the years 1990-1999.} \label{fig1}
\end{figure}

\begin{figure}[ht]
\begin{center}
\leavevmode
\includegraphics[angle=270,width=15cm]{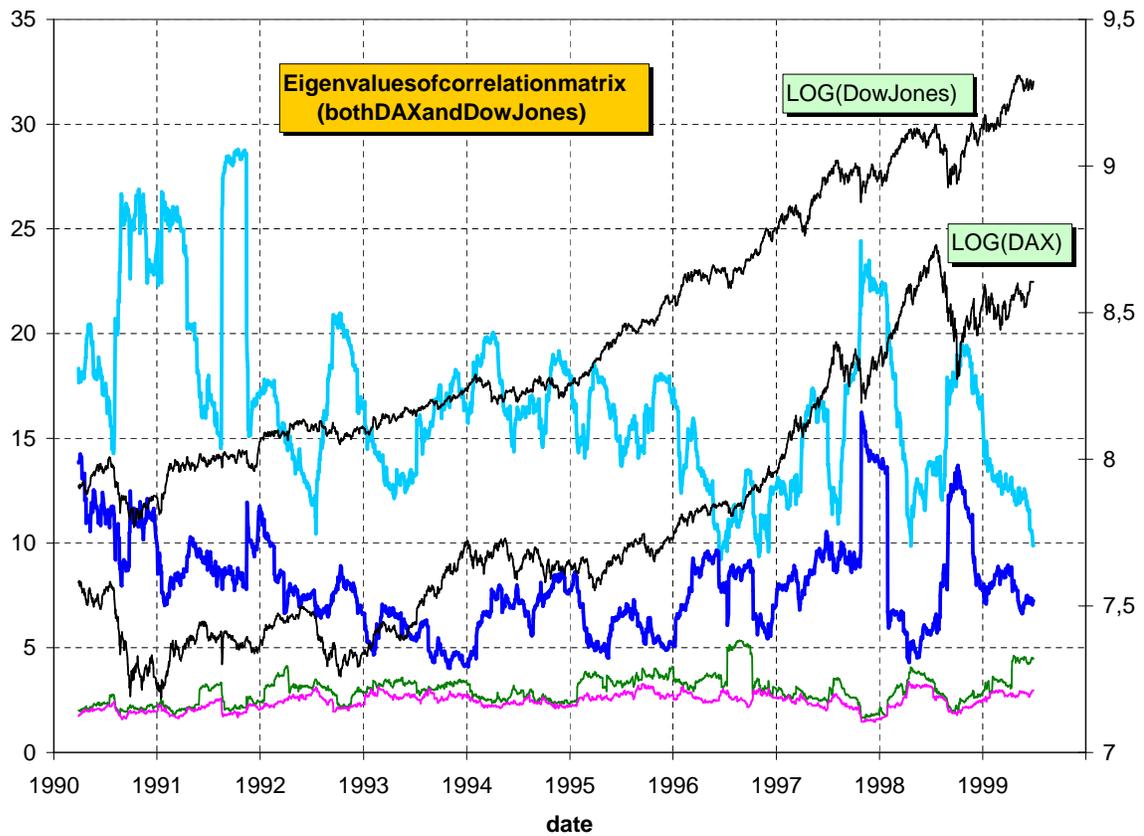}
\end{center}
\caption{\it Time-dependence of four largest eigenvalues corresponding to
the global (DAX + DJ) correlation matrix ${\bf C}_{\cal G}$ calculated 
from the time-series of daily price changes in the interval of $T=60$ past days, 
during the years 1990-1999.
The DAX and the Dow Jones time-variations (represented by their logarithms) 
during the same period are also displayed.} \label{fig2}
\end{figure}

\begin{figure}[ht]
\begin{center}
\leavevmode
\includegraphics[angle=270,width=10cm]{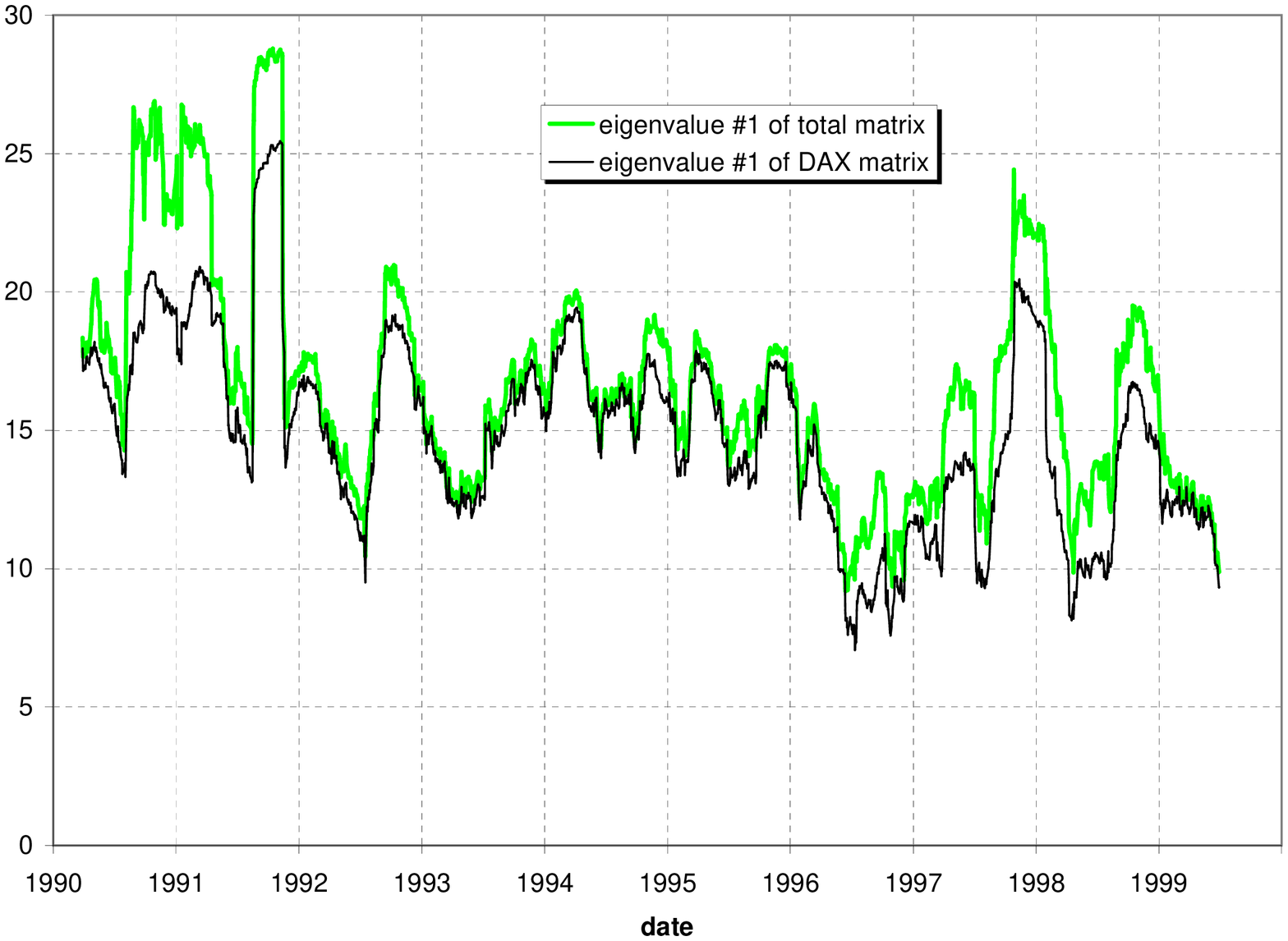}
\includegraphics[angle=270,width=10cm]{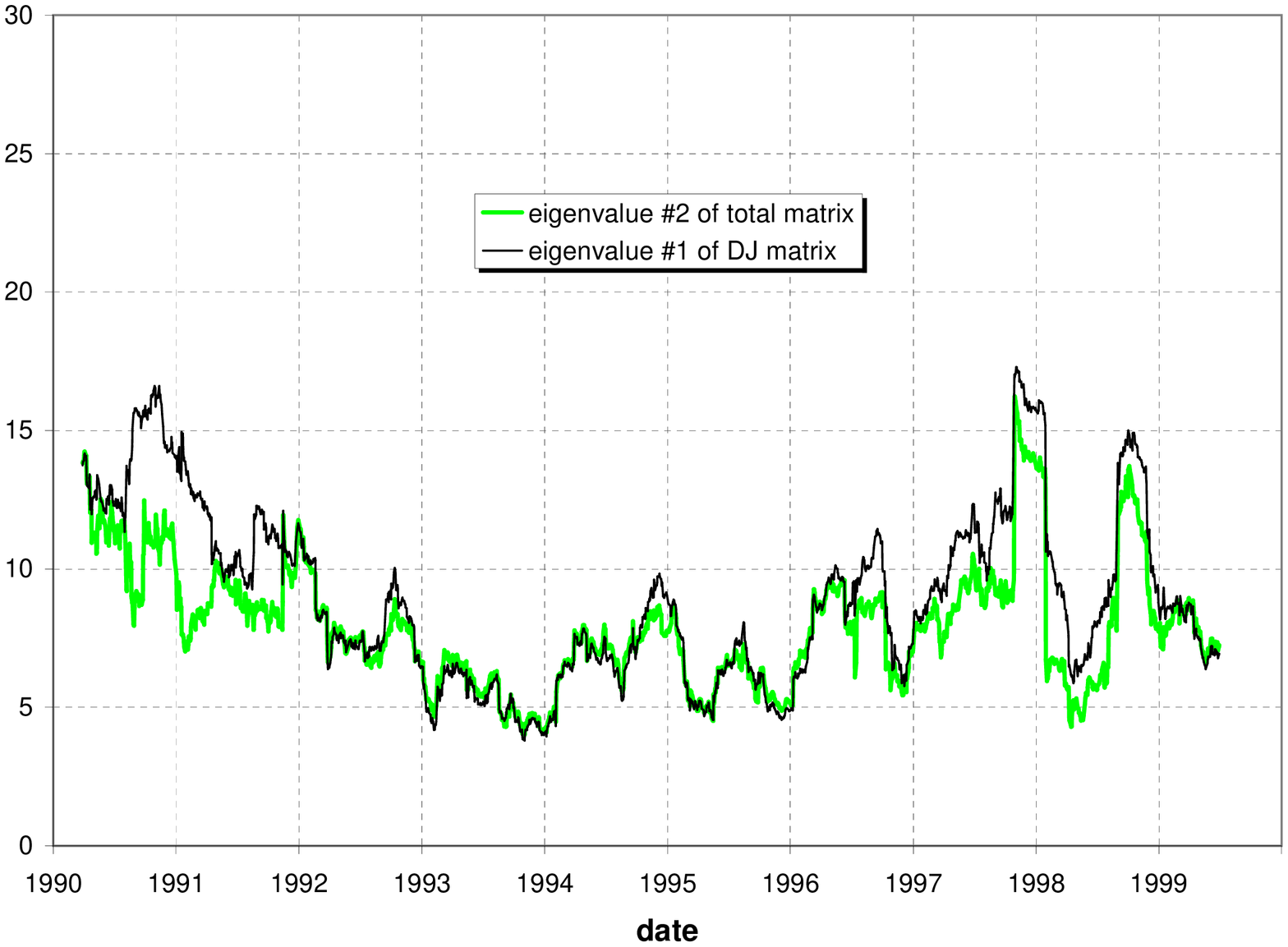}
\end{center}
\caption{\it a) Time-dependence
of the largest eigenvalue $\lambda_{\cal G}^1$ of ${\bf C}_{\cal G}$
(grey thick line) versus 
the largest eigenvalue $\lambda_{DAX}^1$ of ${\bf C}_{DAX,DAX}$ (black thin line).
b) Time-dependence of the second largest eigenvalue 
$\lambda_{\cal G}^2$ of ${\bf C}_{\cal G}$ (grey thick line)
and of the largest eigenvalue $\lambda_{DJ}^1$ of ${\bf C}_{DJ,DJ}$ 
(black thin line). In all these calculations the time window $T$ of the
same length of 60 trading days is consistently used.} \label{fig3}
\end{figure}

\begin{figure}[ht]
\begin{center}
\leavevmode
\includegraphics[angle=270,width=15cm]{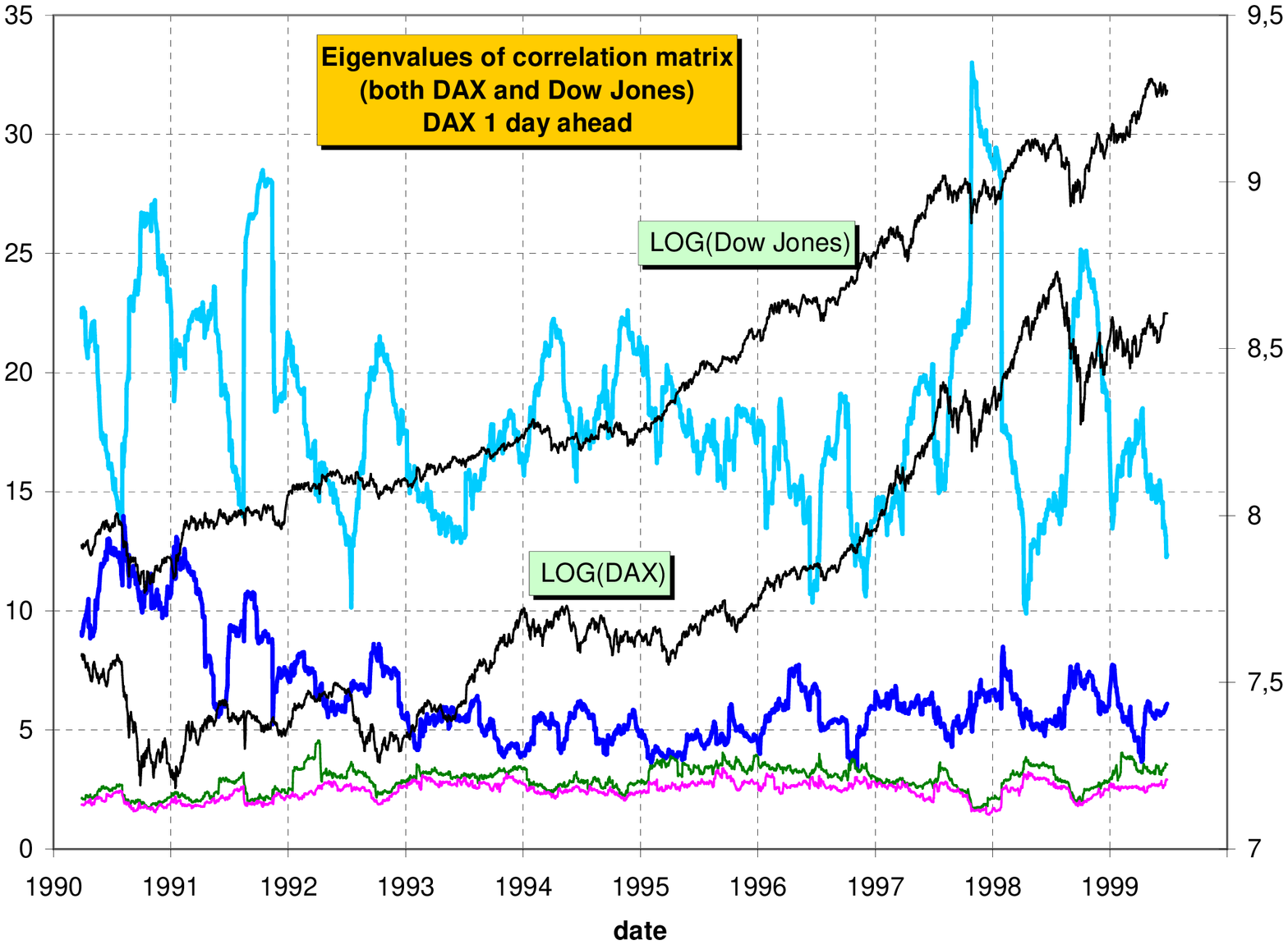}
\end{center}
\caption{\it Same as Fig.~2 but now the DAX returns are shifted one day
ahead relative to the Dow Jones returns when ${\bf C}_{\cal G}$
is constructed.} \label{fig4}
\end{figure}

\begin{figure}[ht]
\begin{center}
\leavevmode
\includegraphics[angle=270,width=15cm]{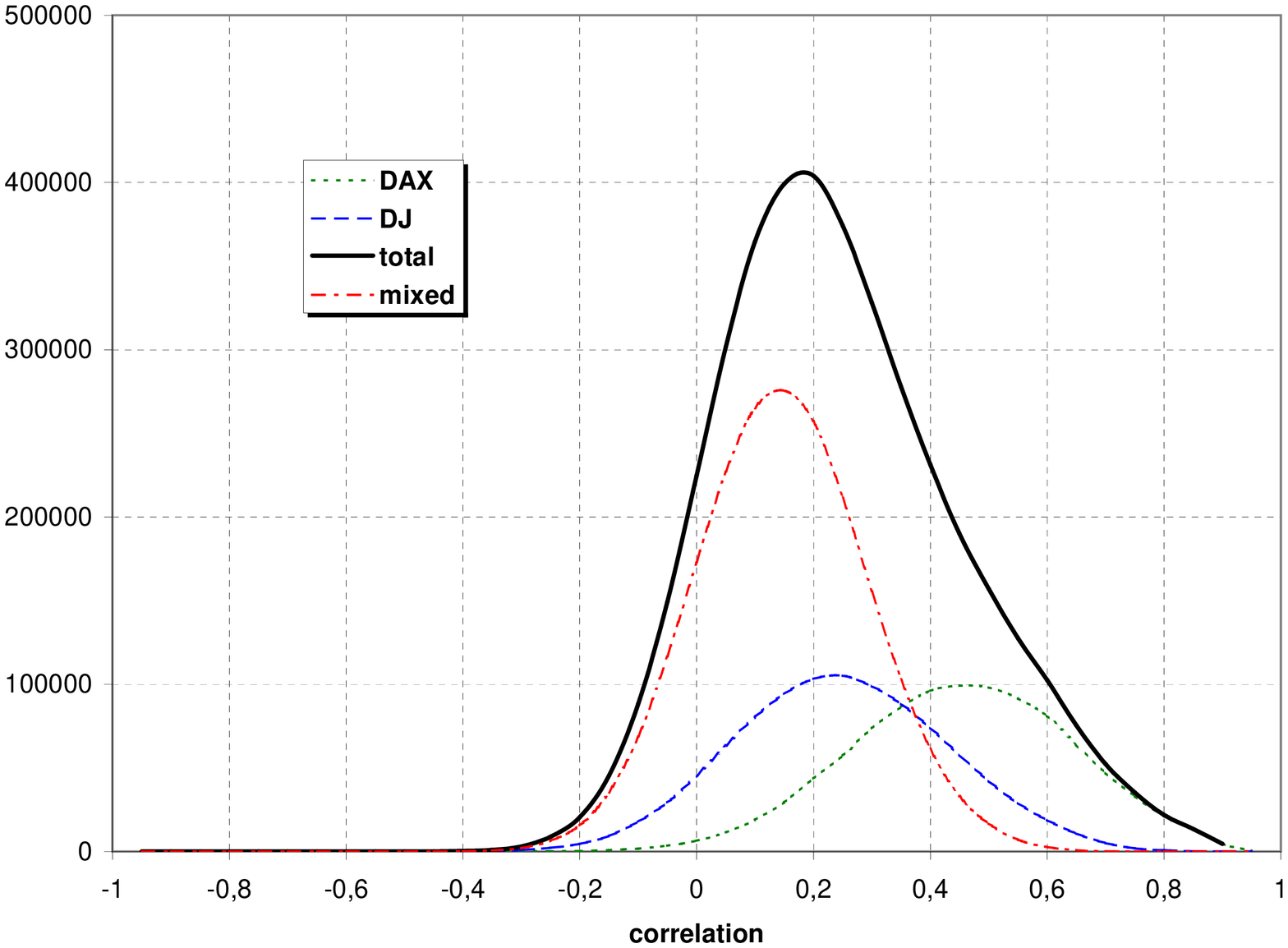}
\end{center}
\caption{\it Same as Fig.~1 but now the DAX returns are shifted one day
ahead relative to the Dow Jones returns when ${\bf C}_{\cal G}$ is constructed.
} \label{fig5}
\end{figure}

\begin{figure}[ht]
\begin{center}
\leavevmode
\includegraphics[angle=270,width=15cm]{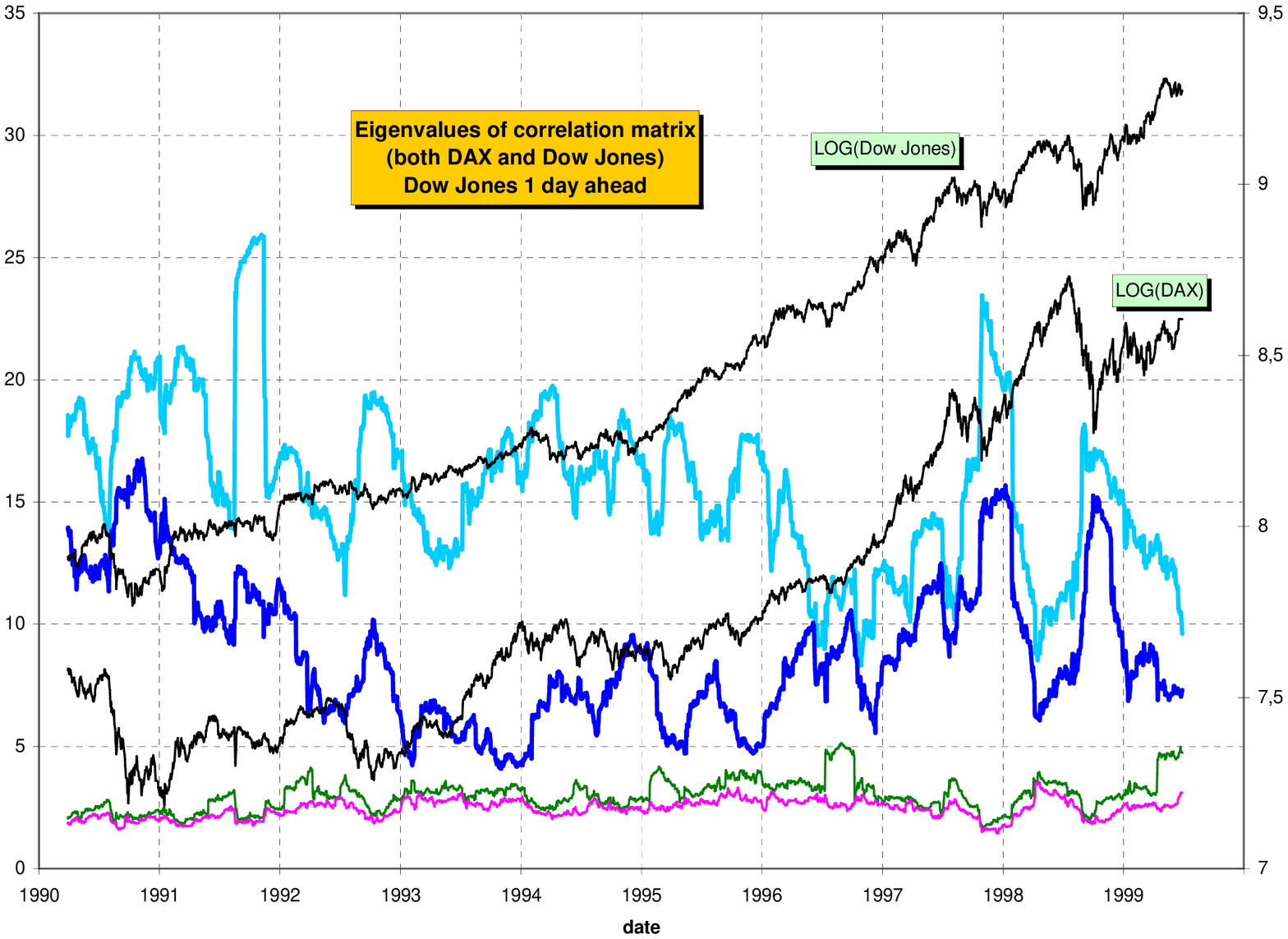}
\end{center}
\caption{\it Same as Fig.~2 but now the Dow Jones returns are shifted one day
ahead relative to the DAX returns when ${\bf C}_{\cal G}$ is constructed.
} \label{fig6}
\end{figure}

\begin{figure}[ht]
\begin{center}
\leavevmode
\includegraphics[angle=270,width=15cm]{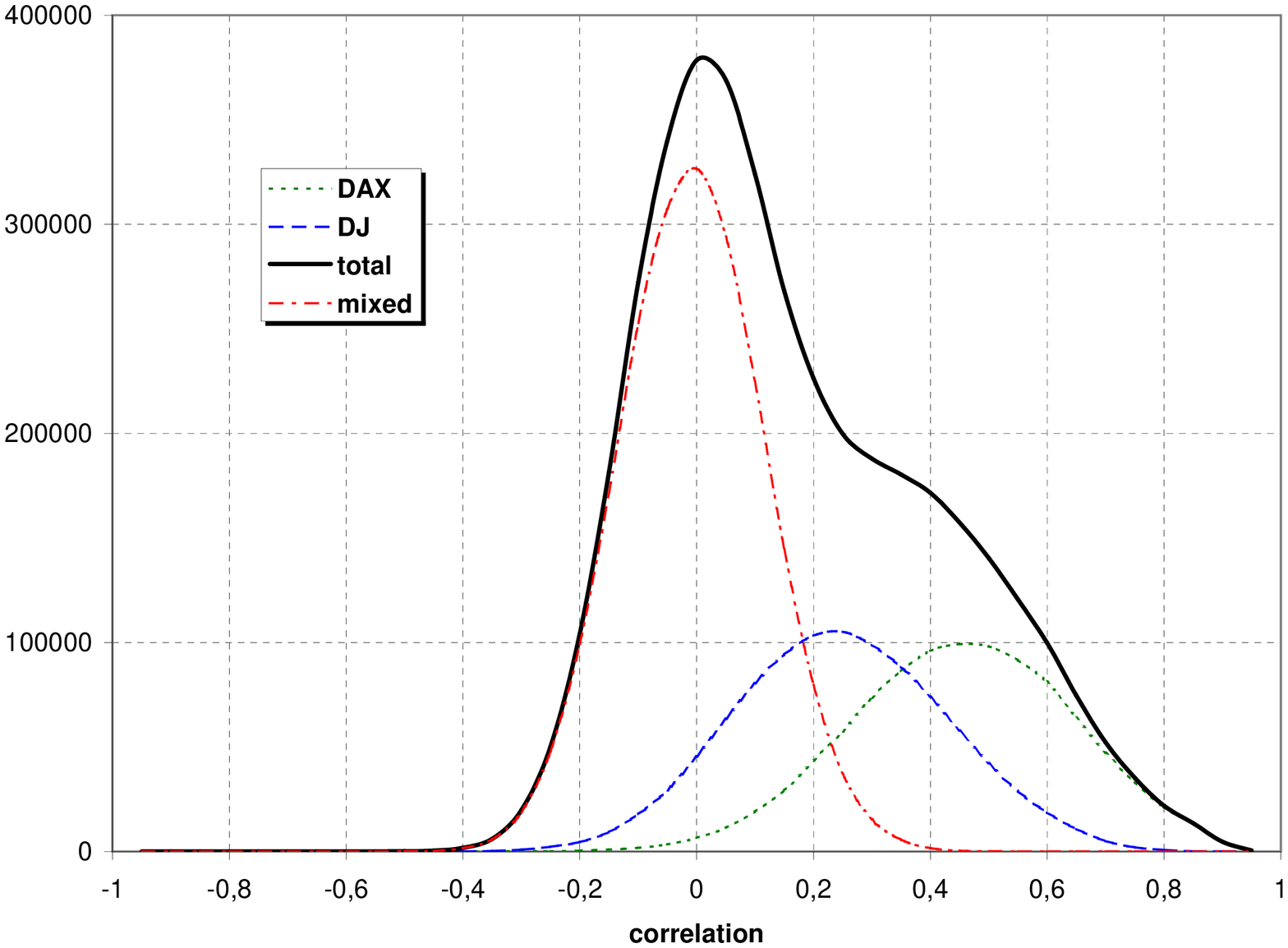}
\end{center}
\caption{\it Same as Fig.~1 but now the Dow Jones returns are shifted one day
ahead relative to the DAX returns when ${\bf C}_{\cal G}$ is constructed.
} \label{fig7}
\end{figure}
\end{document}